# Mössbauer and magnetic measurements of superconducting LiFeP

A. Błachowski[1], K. Ruebenbauer[1*], J. Żukrowski[2], J. Przewoźnik[2], and J. Marzec[3]

[1]Mössbauer Spectroscopy Division, Institute of Physics, Pedagogical University
*PL-30-084 Kraków, ul. Podchorążych 2, Poland*

[2]Solid State Physics Department, Faculty of Physics and Applied Computer Science, AGH University of Science and Technology
*PL-30-059 Kraków, Al. Mickiewicza 30, Poland*

[3]Department of Hydrogen Energy, Faculty of Energy and Fuels, AGH University of Science and Technology
*PL-30-059 Kraków, Al. Mickiewicza 30, Poland*

[*]**Corresponding author:** sfrueben@cyf-kr.edu.pl

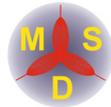

**Abstract**

The LiFeP sample has been prepared by means of high-temperature solid state reaction. A transition temperature to the superconducting state was found as about 5.5 K by measurements of magnetization versus temperature. The second critical field was found above 1 kOe. Mössbauer spectra of the LiFeP phase are characterized by the quadrupole split doublet having splitting of about 0.10 mm/s at room temperature and 0.12 mm/s at 4.2 K. No magnetic order occurs down to 4.2 K.

## 1. Introduction

The compound LiFeP belongs to the "111" group of iron-based superconductors. A transition to the superconducting state was discovered by Deng *et al.* [1] for this compound at about 6 K. Mydeen *et al.* [2] found negative coefficient for the transition temperature versus pressure, i.e., a transition temperature is getting lower upon applying high pressure as for LiFeAs [3]. Shein and Ivanovskii [4] performed *ab inito* calculations of the electronic structure in LiFeP. Boyanov *et al.* [5] investigated iron phosphides doped with lithium and obtained Mössbauer spectrum of LiFeP at room temperature as the minor phase in their sample.

Present contribution is aimed at the preparation of the LiFeP compound, magnetic measurements confirming superconductivity in this system, and Mössbauer measurements – particularly at low temperatures. Mössbauer data were never previously reported for LiFeP at low temperatures to our best knowledge.

## 2. Experimental

The LiFeP sample was obtained by a high temperature solid-state reaction of stoichiometric amounts of Fe and P powders (Alfa Aesar, 99.5 % and 99 %, respectively) and lithium chunks (Aldrich, 99.9 %) used with some excess (1.5 at. %). The reagents were weighed out, initially mixed and placed in a gas tight steel vial (8001 Spex) in an argon filled glove box ($pO_2$ and $pH_2O$ < 0.1 ppm), and then ball milled three times for 5 minutes. After every milling course the vial was opened in the glove box for manual loosening of the batch. Such obtained powder was pressed with 1 GPa under protective, high purity argon atmosphere to obtain pellet 3 mm thick and of diameter 10 mm. The pellet was then wrapped with Au foil and sealed in a silica tube filled with argon-hydrogen mixture (5 % $H_2$ in Ar) at the pressure of a few Tr. The ampoule was placed in a tube furnace and heated initially for 3 hours at 180 $^o$C followed by final heating performed at 800 $^o$C for 20 h and finished with a slow cool down with the furnace. This procedure was developed to avoid any oxidation. Namely, the material was never exposed to the air. The colour of the sample was black/dark grey.

Powder X-ray diffraction pattern was obtained at room temperature by using D5000 Siemens diffractometer. The Cu-K$\alpha_{1,2}$ radiation was used with the pyrolitic graphite monochromator on the detector side. Data were analyzed by the Rietveld method as implemented within the FULLPROF program.

Magnetic measurements were performed by the vibrating sample magnetometer (VSM) of the Quantum Design PPMS-9 system. The sample mass was chosen as 43.837 mg.

Mössbauer spectra were obtained by means of the MsAa-3 spectrometer with the commercial $^{57}$Co(Rh) source kept at room temperature. Absorber was prepared by mixing 30 mg/cm$^2$ of LiFeP sample with the $B_4C$ carrier. Spectra were obtained at room temperature, 77 K and 4.2 K. Data was processed within transmission integral approximation as implemented in the MOSGRAF system. All shifts are reported versus room temperature α-Fe.

## 3. Results and Discussion

Figure 1 shows X-ray diffraction pattern. The sample consists of 93.4(1.5) wt. % of LiFeP of the space group *P4/nmm*. The lattice parameters were found as *a*=3.698(1) Å and $c=6.030(2)$ Å. The fractional co-ordinates along tetragonal axis were determined as follows: $z_{Li}=0.853(5)$ and $z_P=0.226(1)$. Values of the above parameters are similar to values found by Deng *et al.* [1]. The sample is contaminated by 5.2(1.2) wt. % of FeP and by 1.4(0.4) wt. % of $Fe_2P$.

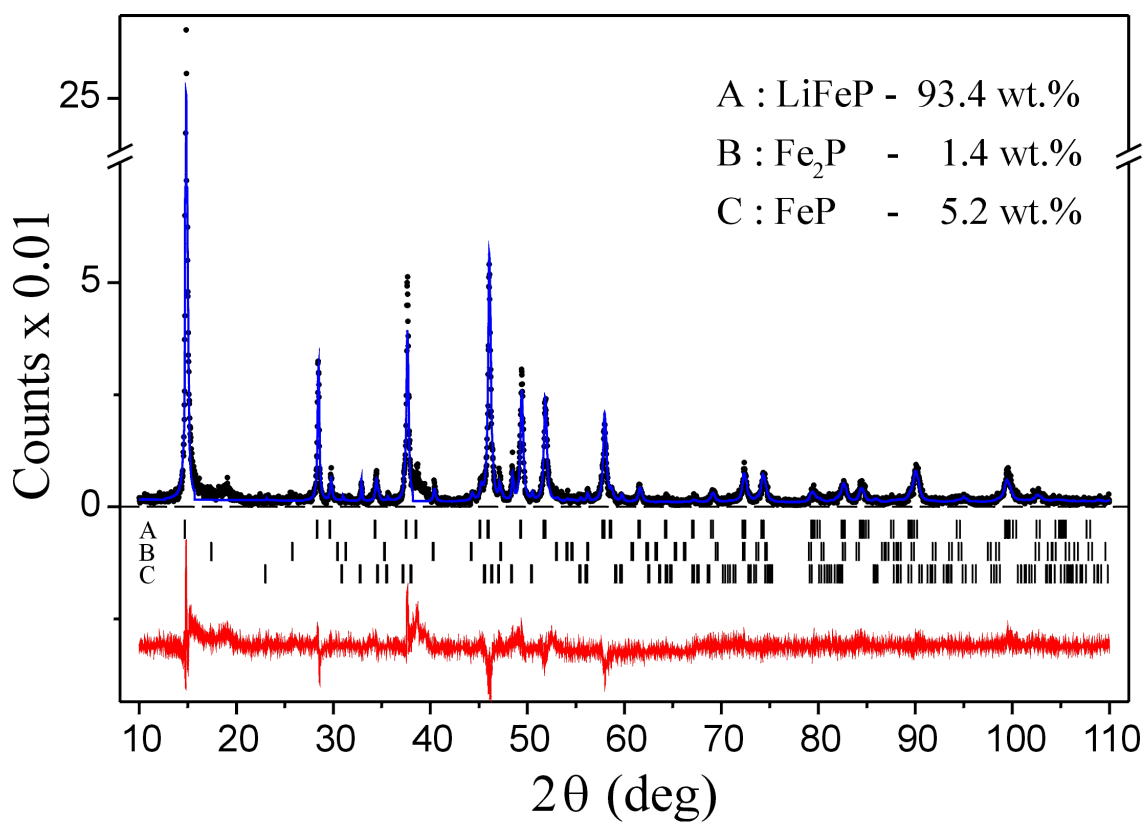

**Figure 1** Room temperature powder X-ray diffraction pattern obtained by using Cu-K$\alpha_{1,2}$ radiation.

Figure 2 shows magnetization versus temperature obtained in the zero field cooling (ZFC) mode for various applied fields. Measurements were performed for 10 Oe and 50 Oe, afterwards hysteresis loops shown in Figure 3 were measured, and subsequently remaining fields (shown in inset of Figure 2) were applied. One can clearly see onset of superconductivity at about 5.5 K for low fields with superconductivity present up to the applied field exceeding 1 kOe. Positive magnetization above transition to the superconducting state is caused by minority phases ordered magnetically at low temperatures. Hysteresis loops were measured at 2 K (below transition to the superconducting state) and at 20 K (above transition). Loops are governed by the minority phases ordered magnetically at both of above temperatures. The low temperature loop shows shoulders due to the presence of superconductor for lower fields. Position of these shoulders is consistent with results of Figure 2 and results of Ref. [2]. Figure 4 shows magnetization obtained in the sweep mode upon cooling from room temperature in the zero field and subsequent warming in the field of 10 kOe (inset). Line shown in red represents magnetization obtained upon cooling in the field of 10 Oe. The lower temperature kink is due to the magnetic ordering of FeP (115 K [6] – complex magnetic structure), while the higher temperature kink corresponds to the magnetic ordering of $Fe_2P$ (215 K [7, 8] – ferromagnetic).

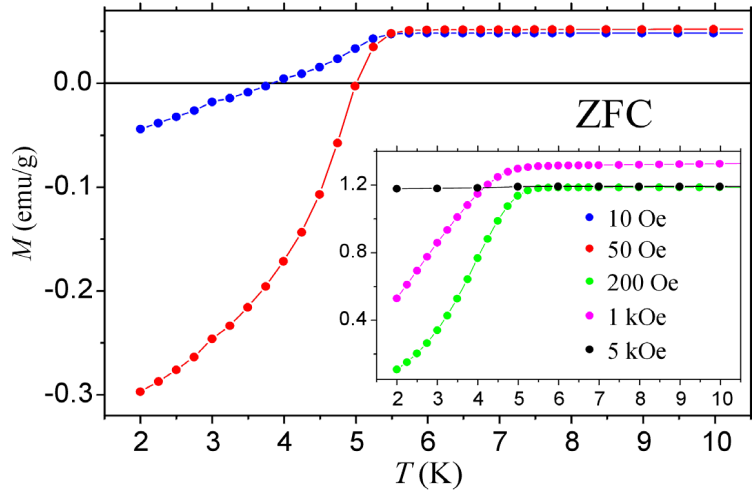

**Figure 2** Magnetization plotted versus temperature for various applied fields.

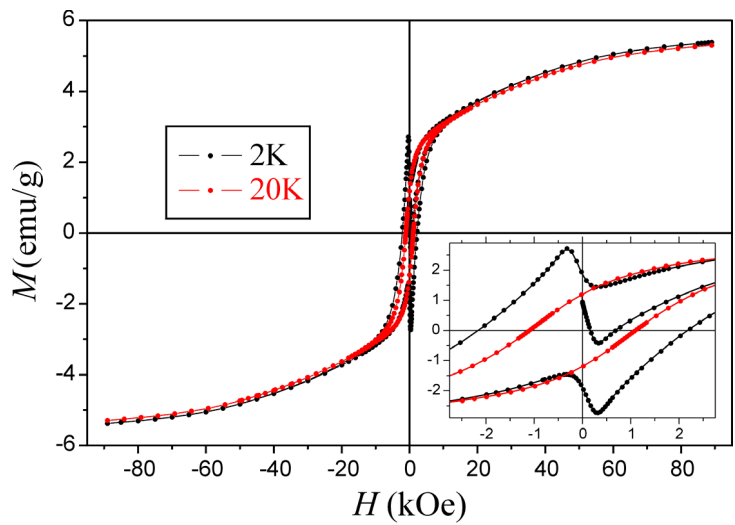

**Figure 3** Hysteresis loops below and above transition to the superconducting state. Inset shows expanded scale of the applied field around zero-field.

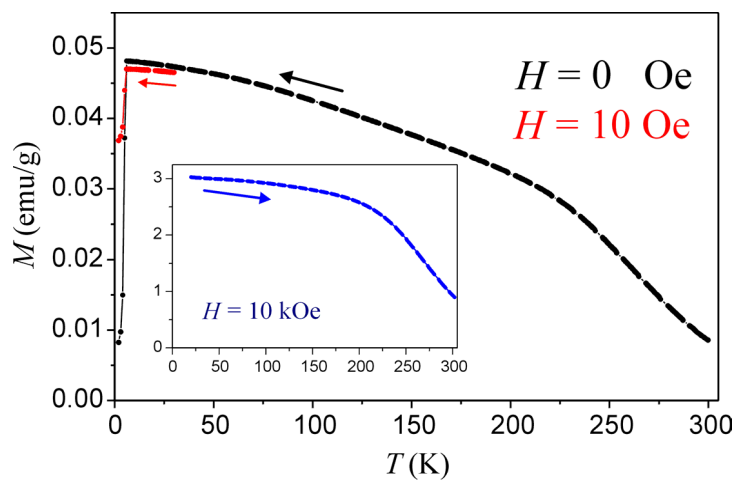

**Figure 4** Magnetization measured versus temperature in various fields.

Mössbauer spectra are shown in Figure 5. LiFeP is described at all temperatures by unresolved quadrupole doublet with the room temperature parameters very close to parameters obtained in Ref. [5]. Relevant parameters are summarized in Table 1. The evolution of the total shift $S$ versus temperature is mainly due to the second order Doppler shift (SOD). The quadrupole splitting $\Delta$ increases with lowering of the temperature, but the change is small and probably due to the thermal redistribution of electrons and non-negligible lattice expansion coefficient at higher temperatures. Lines are generally broad (reported line width is absorber line width – transmission integral). The broadening is enhanced at low temperatures (77 K and 4.2 K) due to the random and non-stoichiometric distribution of lithium ions. Line narrowing at room temperature is hard to explain. It could be either due to the increased lithium mobility or due to the phase transition between room temperature and 77 K as frequently observed in the iron-based superconductors. No magnetic order of LiFeP is found down to 4.2 K. Parameters of the minor phases were taken from literature, i.e., for FeP from Ref. [6] and for $Fe_2P$ from Refs [7, 8]. The average contribution to the spectrum area due to iron in FeP amounts to 5.8(3) %, while the same contribution due to the $Fe_2P$ is 3.7(2) % upon averaging over all temperatures used to collect spectra.

It seems that iron does not bear any magnetic moment in LiFeP in similarity to other iron-based superconductors like "11" FeSe [9], and "1111" compounds [10, 11], where Mössbauer spectra were obtained in the strong external magnetic field at low temperatures and hyperfine field was equal to applied field.

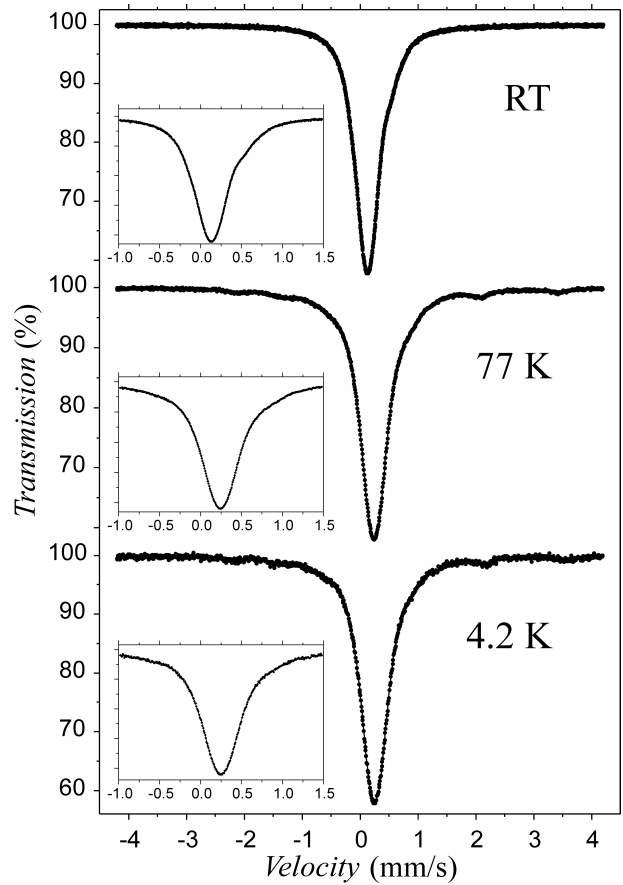

**Figure 5** Mössbauer spectra at various temperatures. Insets show expanded velocity scale around origin, where one finds signal from LiFeP.

**Table 1**

Essential Mössbauer parameters of the LiFeP phase. The symbol $S$ stands for the total shift versus room temperature α-Fe, the symbol $\Delta$ denotes quadrupole splitting, while the symbol $\Gamma$ stands for the absorber line width obtained within transmission integral algorithm.

| $T$ (K) | $S$ (mm/s) | $\Delta$ (mm/s) | $\Gamma$ (mm/s) |
|---|---|---|---|
| RT  | 0.247(1) | 0.101(1) | 0.172(1) |
| 77  | 0.356(1) | 0.112(2) | 0.224(1) |
| 4.2 | 0.364(1) | 0.119(3) | 0.227(2) |